\documentclass[%
 reprint,
superscriptaddress,
 amsmath,amssymb,
 aps,
nomerge
]{revtex4-1}
\usepackage{graphicx}
\usepackage{dcolumn}
\usepackage{bm}
\setcitestyle{numbers,square,citesep={,\kern-.27em}}
\usepackage{multirow}
\begin{document}
\preprint{APS/123-QED}

\title{Intrinsic Charge Transport in Stanene: \\ Roles of Bucklings and Electron-Phonon Couplings}


 \author{Yuma Nakamura}
  \thanks{These two authors contributed equally}
 \affiliation{MOE Key Laboratory of Organic OptoElectronics and Molecular Engineering,
  Department of Chemistry, Tsinghua University, Beijing 100084, P. R. China}
 \affiliation{Institute for Materials Research, Tohoku University, Sendai, 980-8577, Japan}
 \author{Tianqi Zhao}
  \thanks{These two authors contributed equally}
 \affiliation{MOE Key Laboratory of Organic OptoElectronics and Molecular Engineering,
  Department of Chemistry, Tsinghua University, Beijing 100084, P. R. China}
 \author{Jinyang Xi}
 \affiliation{Materials Genome Institute, Shanghai University, Shanghai, 200444, P. R. China}
 \author{Wen Shi}
 \affiliation{MOE Key Laboratory of Organic OptoElectronics and Molecular Engineering,
  Department of Chemistry, Tsinghua University, Beijing 100084, P. R. China}
  \author{Dong Wang}
  \email{Corresponding author: dong913@tsinghua.edu.cn}
  \affiliation{Materials Genome Institute, Shanghai University, Shanghai, 200444, P. R. China}
  \author{Zhigang Shuai}
  \email{Corresponding author: zgshuai@tsinghua.edu.cn}
  \affiliation{MOE Key Laboratory of Organic OptoElectronics and Molecular Engineering,
  Department of Chemistry, Tsinghua University, Beijing 100084, P. R. China}
  \affiliation{Key Laboratory of Organic Solids, Beijing National Laboratory for Molecular Science (BNLMS), Institute of Chemistry,\\Chinese Academy of Science, Beijing, 100190, P. R. China}
  \affiliation{Collaborative Innovation Center of Chemistry for Energy Materials, Xiamen University, Xiamen, 351005, P. R. China}



\begin{abstract}
The intrinsic charge transport of stanene is investigated by using density functional theory and density functional perturbation theory coupled with Boltzmann transport equations at the first-principles level. The Wannier interpolation scheme is applied to calculate the charge carrier scatterings with all branches of phonons considering dispersion for the whole range of the first Brillouin zone. The intrinsic electron and hole mobilities are calculated to be 2-3 $\times$ 10$^3$ cm$^2$ V$^{-1}$ s$^{-1}$ at 300 K. It is found that the intervalley scatterings from the out-of-plane and the transverse acoustic phonon modes dominate the carrier transport process. By contrast, the mobilities obtained by the conventional deformation potential approach are found to be as large as 2-3 $\times$ 10$^6$ cm$^2$ V$^{-1}$ s$^{-1}$ at 300 K, in which the longitudinal acoustic phonon scattering in the long wavelength limit is assumed to be the dominant scattering mechanism. The inadequacy of the deformation potential approximation in stanene is attributed to the buckling in its honeycomb structure, which originates from the sp$^2$-sp$^3$ orbital hybridization and breaks the planar symmetry. This paper further proposes a strategy to enhance carrier mobilities by suppressing the out-of-plane vibrations through clamping by a substrate. 
\end{abstract}

\pacs{Valid PACS appear here}
\maketitle


\section{\label{sec:level1}Introduction}
\frenchspacing
Motivated by the peculiar behavior of electronic structure with Dirac cone in graphene,\cite{Novoselov2004ElectricFilms,Geim2007TheGraphene,Novoselov2005Two-dimensionalGraphene,Li2008MATERIALSMaterials} the class of 2D materials have received intensive interests in recent years.\cite{Gupta2015RecentGraphene} The group IV elemental analogues of graphene, including silicene,\cite{Vogt2012Silicene:Silicon} germanene,\cite{Li2014BuckledPt111} and stanene,\cite{Balendhran2015ElementalPhosphorene,Xu2013Large-GapFilms,Matthes2013MassiveTinene.,Zhu2015EpitaxialStanene} are promising alternatives going beyond graphene owing to their outstanding electronic properties. Molecular beam epitaxy technique facilitates the materials fabrications.\cite{Vogt2012Silicene:Silicon,Li2014BuckledPt111,Zhu2015EpitaxialStanene} Among these group IV elemental 2D sheets, stanene draws particular attention since it has been predicted to be a topological insulator with large band gap theoretically,\cite{Kane2005QuantumGraphene,Qi2011TopologicalSuperconductors,Hasan2010Colloquium:Insulators,Yao2007Spin-orbitCalculations} arousing interests in dissipationless electronics. Moreover, stanene also shows enhanced thermoelectric performance\cite{Xu2014EnhancedInsulators,Peng2016LowStanene} near-room-temperature quantum anomalous effect,\cite{Yu2010QuantizedInsulators} and topological superconductivity.\cite{Shaidu2016FirstStanene,Ezawa2015MonolayerStanene} For these applications, the carrier transport lies in the center of electronic processes. Electron-phonon (el-ph) couplings play an essential role in determining the intrinsic transport properties. For evaluating the intrinsic carrier mobility of 2D materials, Long et al.\cite{Long2009TheoreticalGraphene,Long2011ElectronicPredictions} have first applied the deformation potential approximation (DPA) proposed by Shockley and Bardeen\cite{Bardeen1950DeformationCrystals} as a first-principles method, where only the scattering by longitudinal acoustic (LA) phonons are taken into account. This approach is conceptually simple but has been extremely successful as has been widely applied in calculating the intrinsic carrier mobility of 2D sheets or nanoribbons, such as graphene,\cite{Xi2012First-principlesNanomaterials,Xi2014Electron-phononApproach,Borysenko2010First-principlesGraphene,Hwang2008AcousticGraphene,Kaasbjerg2012UnravelingGraphene} silicene,\cite{Shao2013First-principlesSilicene} germanene,\cite{Ye2014IntrinsicCalculations} graphdiyne,\cite{Long2011ElectronicPredictions,Xi2012First-principlesNanomaterials} $\alpha$-graphyne,\cite{Chen2013CarrierGraphene.pdf} transition metal dichalcogenide (TMD),\cite{Cai2014Polarity-reversedNanoribbons} phosphorene,\cite{Qiao2014High-mobilityPhosphorus,Fei2014EnhancedPhosphorene} as well as perovskites.\cite{Zhao2016IntrinsicFirst-Principles}

DPA can overestimate the intrinsic room-temperature mobility because it only considers the longitudinal acoustic phonon scattering process. Taking the case of silicene as an example, DPA predicts the intrinsic mobility to be 2$\times$10$^5$ cm$^2$ V$^{-1}$ s$^{-1}$.\cite{Ye2014IntrinsicCalculations} By contrast, when electron scatterings with all branches of phonons were accounted by using the density functional perturbation theory (DFPT), the mobility of silicene becomes 2100 cm$^2$ V$^{-1}$ s$^{-1}$,\cite{Gunst2016First-principlesMaterials} or even smaller value is also reported (1200 \cite{Li2013IntrinsicPrinciples} and 750 \cite{Gaddemane2016TheoreticalSimulations}  cm$^2$ V$^{-1}$ s$^{-1}$). This comes from the fact that the reduced symmetry of silicene (D$_{3d}$) compared to graphene (D$_{6h}$) causes additional scattering processes with phonons other than LA.\cite{Gaddemane2016TheoreticalSimulations,Fischetti2016Mermin-WagnerSymmetry}

It is thus intriguing to understand how and when DPA fails. In fact, according to Shockley and Bardeen's original argument, the room-temperature (300 K) accounts for about a de Broglie wavelength of 7 nm for electron, which is much longer than lattice spacing. Thus, electron scatterings with acoustic phonons have been believed to be the dominant process.\cite{Bardeen1950DeformationCrystals} Carrier transports in iodine-functionalized stanene nanoribbons have been studied by using Kubo-Greenwood formalism, where el-ph couplings are approximated by calculating deformation potentials for longitudinal and transverse acoustic phonons.\cite{Vandenberghe2014CalculationNanoribbons} In this work, we consider all the el-ph scattering processes in stanene to calculate the carrier mobility. The state-of-the-art DFPT\cite{Baroni2001PhononsTheory} coupled with a Wannier interpolation scheme\cite{Marzari2012MaximallyApplications} as implemented in the Quantum ESPRESSO,\cite{Giannozzi2009QUANTUMMaterials.} Wannier90,\cite{Mostofi2014AnFunctions} and EPW packages\cite{Noffsinger2010EPW:Functions,Ponce2016EPW:Functionsb} was employed to obtain the ultradense electronic band structures, phonon dispersion, and el-ph coupling matrix elements. With all these ingredients, the Boltzmann transport theory with relaxation time approximation was used to determine the intrinsic charge carrier mobilities. This approach was conducted to find the limitation of DPA. Buckling in the hexagonal honeycomb structure of stanene, originating from the sp$^2$-sp$^3$ orbital hybridization\cite{Cahangirov2009Two-Germanium,Liu2011Low-energyTin,Jose2012UnderstandingSilicene} can lead to remarkable difference from the planar structures such as graphene or graphynes, where huge intrinsic mobility  $\mu \sim$ 10$^5$ cm$^2$ V$^{-1}$ s$^{-1}$ has been predicted.\cite{Long2011ElectronicPredictions} We note that the mobility in nonplanar structured monolayer MoS$_2$ was found to be 150-410 cm$^2$ V$^{-1}$ s$^{-1}$, \cite{Cai2014Polarity-reversedNanoribbons,Li2013IntrinsicPrinciples,Kaasbjerg2012Phonon-limitedPrinciples,Restrepo2014AMaterials,Li2015ElectricalMoS2} and phosphorene to be 170-460 cm$^2$ V$^{-1}$ s$^{-1}$.\cite{Liao2015AbPhosphorene,Jin2016HighlyPrinciples} It is thus intriguing to look at the contribution of each phonon mode to carrier transport and to compare with DPA for the buckled stanene layer.

\section{\label{sec:level1}Results and Discussion}
\subsection{Electronic Structures and Phonon Dispersions}
The crystal structure of stanene is illustrated in Fig. 1(a). The optimized lattice constant is 4.676 {\AA}, which is in good agreement with previous calculations (4.676 \cite{Xu2013Large-GapFilms} or 4.673 \cite{Matthes2013MassiveTinene.}  {\AA}) under the generalized-gradient approximation. The lattice constant matches with that of Bi$_2$Te$_3$ substrate (4.383 {\AA}).\cite{Zhu2015EpitaxialStanene} The obtained value of buckling 0.85 {\AA} also agrees with previous calculation.\cite{Matthes2013MassiveTinene.} The experimental buckling value of stanene on Bi$_2$Te$_3$ substrate was found to be 1.2 $\pm$ 0.2 {\AA}, which was ascribed to strain-induced enhancement effect.\cite{Zhu2015EpitaxialStanene} The unit cell of stanene consists of two sublattices and the reciprocal space is shown in Fig.1(b) with the first Brillouin zone and the high symmetry points $\boldsymbol{\Gamma}=(0, 0)$, $\mathbf{K}=(1/3, 1/3)$, $\mathbf{K'}=(2/3, -1/3)$, and $\mathbf{M}=(1/2, 0)$ whose base is spanned by $\mathbf{b}_1$ and $\mathbf{b}_2$. The Dirac cones are seen at $\mathbf{k = K}$ and $\mathbf{k = K'}$, which we will denote as Dirac point {\it K} and {\it K \!}$'$, respectively. The band structures of stanene are depicted in Fig. 1(c) to compare with that of graphene. Similar to graphene, the Dirac cones of stanene appear at $\mathbf{k = K}$ and $\mathbf{k = K'}$ in the first Brillouin zone when spin-orbit coupling (SOC) is not included. However, a band gap of 76 meV is opened when SOC is taken into account (Fig. 1(c)). The band gap is also in agreement with a previous work (73 meV).\cite{Matthes2013MassiveTinene.} This band gap opening indicates the possibility to realize topological insulator phase. We obtain the Fermi velocities of 1.44 $\times$ 10$^6$ and 0.83 $\times$ 10$^6$ m s$^{-1}$ at the Dirac point for graphene and stanene, respectively. The smaller Fermi velocity in stanene is attributed to the less dispersive band structure near the Dirac point.

The phonon dispersion relations of stanene and graphene are shown in Fig. 1(d). The phonon branches are classified as out-of-plane acoustic (ZA), transverse acoustic (TA), LA, out-of-plane optical (ZO), transverse optical (TO), and longitudinal optical (LO). The flexural ZA mode of graphene obey a quadratic dispersion $\omega_{ZA} \propto |q|^2$, which is attributed to the sixfold rotational symmetry.\cite{Cahangirov2009Two-Germanium,Mariani2008FlexuralGraphene,Sahin2009MonolayerCalculations,Ribeiro-Soares2014GroupSystems,Manes2007Symmetry-basedGraphene,Peng2016UnexpectedInitio,Carrete2016PhysicallyBorophene} The obtained phonon frequency of stanene is about ten times smaller than that of graphene because of the heavier atomic mass. The intersection of LA and ZO in graphene is absent in stanene.
\begin{figure}[t]
\centering
    \includegraphics[width=\linewidth]{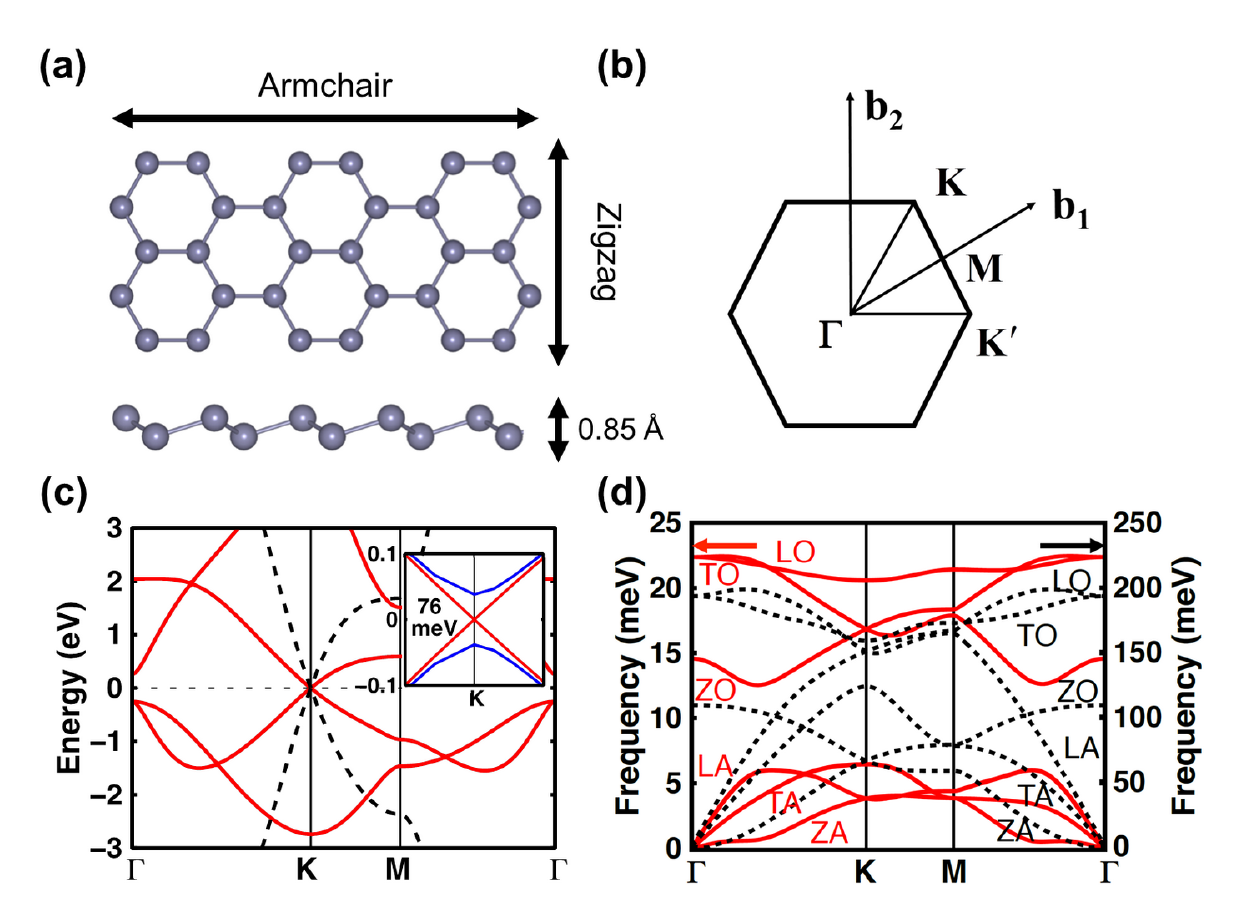}
   \caption{(a) Top and side view of stanene. (b) Schematic illustration of the first Brillouin zone and high symmetry points. (c) Band structures of graphene (black-dashed), stanene without SOC (red-solid) and with SOC (blue-solid, inset). (d) Phonon dispersions of graphene (black-dashed) and stanene (red-solid). }
  \label{fgr:band-dispersion}
\end{figure}

\subsection{Full Electron-Phonon Couplings and Deformation Potentials}

 \begin{table*}[tp]
\small
  \caption{Lattice constant, deformation potential (DP) constant $D_{LA}$, elastic constant $C_{2D}$, and carrier mobility $\mu$ for stanene and graphene along zigzag and armchair dictions at 300 K. DP constants from full evaluation of el-ph coupling are given in parentheses.}
  \label{tbl:example}
  {\renewcommand{\arraystretch}{1.4}
  \begin{tabular*}{1.0\textwidth}{@{\extracolsep{\fill}}cccccccc}
    \hline     \hline
    System & Buckling ({\AA})&  Direction & Lattice Const. ({\AA}) &  $D_{LA}$(eV) & $C_{2d}$ (J m$^{-1}$) &  $\mu_e$ (cm$^2$ V$^{-1}$ s$^{-1}$)  &$\mu_h$ (cm$^2$ V$^{-1}$ s$^{-1}$) \\
   \hline
      \multirow{2}{*}{
          Stanene
        }&     \multirow{2}{*}{
           0.85
        }
         & Zigzag  &4.67&0.48&28.5&2.77 $\times 10^6$&4.01 $\times 10^6$\\ 
     & & Armchair  &8.09 &0.47&28.6&2.44 $\times 10^6$ &3.52 $\times 10^6$  \\

     \multirow{2}{*}{
          Graphene
        }&     \multirow{2}{*}{
           0.00
        }
        & Zigzag  &2.47 &5.14&328&3.32 $\times 10^5$&2.05 $\times 10^5$\\
    & & Armchair  &4.27  &5.00&328&3.32 $\times 10^5$&2.05 $\times 10^5$\\
    \hline      \hline
  \end{tabular*}
  }
\end{table*}

\begin{figure*}[t]
\centering
 \includegraphics[width=\linewidth]{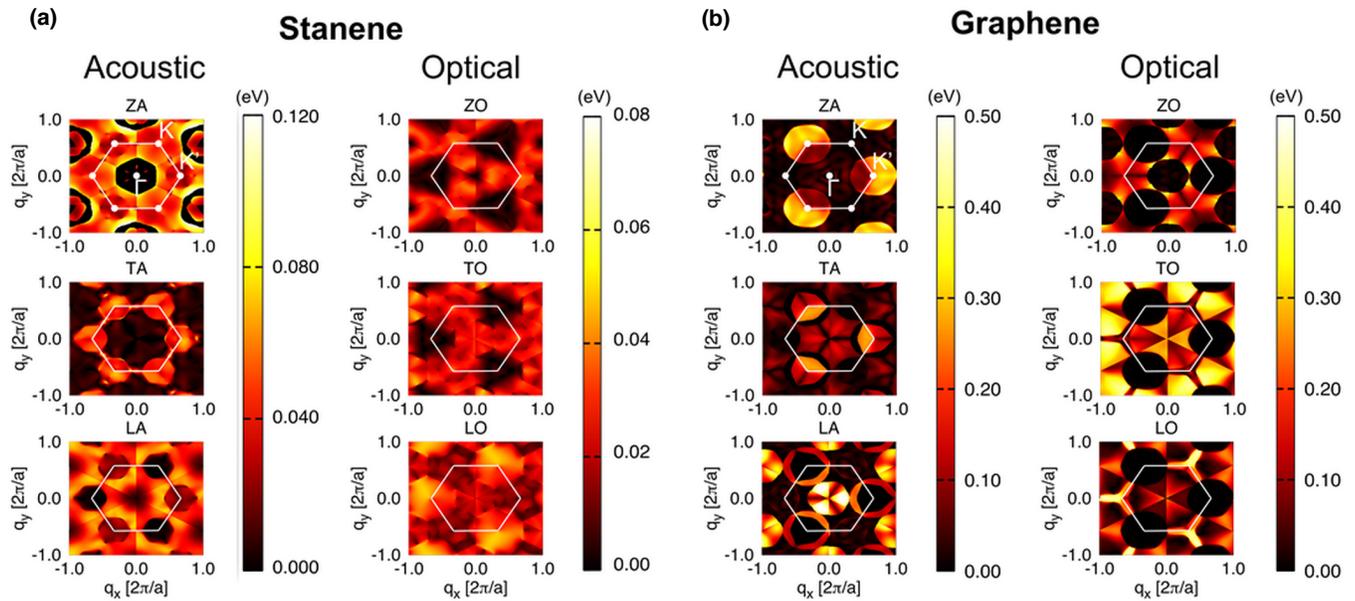}
   \caption{Electron-phonon coupling matrix of (a) stanene and (b) graphene as a function of phonon wavevector $\mathbf{q}$ at a fixed $\mathbf{k}$-point {\it K} in the conduction band. The white line shows first Brillouin zone. The labels $\boldsymbol{\Gamma}$, $\mathbf{K}$, and $\mathbf{K'}$ describe the high symmetry point of phonon wavevector $\mathbf{q}$,which correspond to the final state $\mathbf{k'=K}$, $\mathbf{K'}$, and $\boldsymbol{\Gamma}$, respectively. }
  \label{fgr:el-ph}
\end{figure*}

In order to elucidate the contribution of each phonon modes to carrier scatterings, we depict the absolute value of el-ph coupling matrix elements (see Eq. (3) in the Appendix) for the Dirac point {\it K} in the conduction band of stanene and graphene as a function of phonon wavevector $\mathbf{q}$ in Cartesian coordinate over the first Brillouin zone (Fig. 2). The el-ph couplings for the other Dirac point {\it K \!}$'$ is shown in the Supporting Information. According to the momentum conservation law, $\mathbf{q} = \boldsymbol{\Gamma}$, $\mathbf{K}$, and $\mathbf{K'}$ correspond to the final electronic state with wavevector $\mathbf{k' = K, K'}$, and $\boldsymbol{\Gamma}$, respectively, with a difference of the lattice vector of reciprocal space. Regions further satisfying the energy conservation law $\varepsilon_\mathbf{k'} = \varepsilon_\mathbf{k} \pm \hbar\omega_\mathbf{q}$ may contribute to carrier scattering process. The el-ph couplings around the center of Brillouin zone demonstrate intravalley transitions, while intervalley transitions are shown near the point $\mathbf{q = K}$. For the TA and LA modes in stanene, the intensity of el-ph coupling over the intervalley transition regions are complementary, that is, the regions with high intensity for TA are always low intensity in LA and vice versa. This complementary behavior between TA and LA has also been demonstrated for silicene in intravalley region.\cite{Gunst2016First-principlesMaterials} Remarkably, the high intensity in intervalley region around $\mathbf{q = K}$ for TA mode is not seen in graphene, which is a striking difference for stanene. It is also found that the el-ph coupling strengths for stanene are several times smaller than those for graphene for all phonon modes. This indicates that the phonon vibrations have weaker effect on electronic states in stanene, which agrees with the decreasing trend from graphene to silicene.\cite{Gunst2016First-principlesMaterials}

Near the intravalley region ($\mathbf{q} = \boldsymbol{\Gamma}$) of Fig. 2, the el-ph coupling matrix for LA depends linearly on the length of the phonon wavevector $|\mathbf{q}|$ in long wavelength limit (see Eq. (12) in Appendix). The slope is known as the deformation potential constant $D_{LA}$, which is often used for estimating the strength of el-ph couplings both experimentally and theoretically. $D_{LA}$ can also be obtained by simulating lattice dilation of the unit cell and by measuring the change of Fermi level with respect to strains (Fig. S2, Supporting Information).\cite{Long2009TheoreticalGraphene,Long2011ElectronicPredictions} Table 1 shows that the two methods can give consistent $D_{LA}$ values. For graphene, the values of $D_{LA}$ along zigzag and armchair directions exhibit almost the same values (5.14 and 5.0 eV, repectively), which is also in agreement with previous theoretical results.\cite{Xi2012First-principlesNanomaterials} Stanene also shows nearly the same $D_{LA}$ values along zigzag and armchair directions, but about one order of magnitude smaller compared to graphene. This demonstrates that the el-ph coupling strength for stanene is one order of magnitude weaker than graphene. This is consistent with the trend that the value of $D_{LA}$ decreases with the buckling height. Namely, for graphene, silicene, germanene, and stanene, the buckling height are 0.0, 0.45,\cite{Jose2012UnderstandingSilicene,Ding2013DensityProperties} 0.69,\cite{Balendhran2015ElementalPhosphorene,Matthes2013MassiveTinene.} and 0.85 {\AA} , and the value of $D_{LA}$ are 5.0, 2.13,\cite{Shao2013First-principlesSilicene} 1.16,\cite{Ye2014IntrinsicCalculations} and 0.48 eV, respectively. In addition, the elastic constants $C_{2D}$ show a decreasing trend: 328.30, 85.99,\cite{Shao2013First-principlesSilicene} 55.98,\cite{Ye2014IntrinsicCalculations} and 28.5 J m$^{-1}$, correspondingly. According to the deformation potential theory, : $\mu \propto C_{2D}/D_{LA}^2$, the decreasing trend of $D_{LA}$ is more prominent and may lead to an increasing of mobility $\mu$ . However, the reduced symmetry of stanene ($D_{3d}$) compared to graphene ($D_{6h}$) may cause additional scattering with phonons other than LA.\cite{Cahangirov2009Two-Germanium,Mariani2008FlexuralGraphene,Sahin2009MonolayerCalculations} We account for the further scattering process in addition to LA-phonon-limited scattering in the following section.

  \begin{table}[htbp]
\small
  \caption{\ Scattering rate of each phonon mode for graphene and stanene at Dirac point  {\it K} and $T=300$. }
  \label{tbl:mode-resolved}
    {\renewcommand{\arraystretch}{1.4}
    \begin{tabular*}{0.5235\textwidth}{@{\extracolsep{\fill}}ccccc}
    \hline \hline
          {\small Scattering\! rate\!}
       &  \multicolumn{2}{c}{Stanene}  & \multicolumn{2}{c}{Graphene}  \\
    $1/\tau$ (s$^{-1}$) &Hole & Electron & Hole & Electron \\
    \hline
    ZA & $1.83 \times 10^{12}$   & $1.84 \times 10^{12}$& $3.55 \times 10^7$    & $2.33 \times 10^7$  \\
    TA  & $1.16 \times 10^{12}$   & $1.16 \times 10^{12}$ & $2.63 \times 10^{10}$ & $6.44 \times 10^9$ \\
    LA   & $2.26 \times 10^{10}$ & $2.21 \times 10^{10}$ & $2.74 \times 10^{11}$ & $2.12 \times 10^{11}$ \\
    ZO & $1.28 \times 10^{10}$     & $1.31 \times 10^{10}$ & $9.77 \times 10^9$    & $8.85 \times 10^9$ \\
    TO& $8.23 \times 10^{10}$ & $8.19\times 10^{10}$  & $1.36 \times 10^{10}$ & $1.43 \times 10^{10}$ \\
    LO  & $2.37 \times 10^{11}$ & $2.29 \times 10^{11}$ & $3.99 \times 10^{10}$ & $5.32 \times 10^{10}$ \\
    Total   & $3.35 \times 10^{12}$ & $3.35 \times 10^{12}$& $3.64 \times 10^{11}$ & $2.85 \times 10^{11}$ \\
    \hline \hline
  \end{tabular*}
  }
\end{table}

\subsection{Intrinsic Carrier Mobility}

  \begin{table*}[htbp]
\small
  \caption{\ Scattering rate of each phonon mode for electrons and holes in stanene and graphene at Dirac point  {\it K} and $T=300$. }
  \label{tbl:InterIntra}
  {\renewcommand{\arraystretch}{1.4}
    \begin{tabular*}{1.0\textwidth}{@{\extracolsep{\fill}}ccccccc}
       \hline \hline
          Scattering rate 
       &  \multicolumn{3}{c}{Stanene}  & \multicolumn{3}{c}{Graphene}  \\
    $1/\tau$ (s$^{-1}$) &Intravalley& Intervalley &Total&  Intravalley &Intervalley & Total \\
    \hline
    ZA & $1.27 \times 10^{3}$    & $1.84 \times 10^{12}$ & $1.84 \times 10^{12}$ &  $1.15 \times 10^{-4}$&$2.33 \times 10^{7}$ &$2.33 \times 10^{7}$ \\
    TA  &$9.11 \times 10^{7}$      & $1.16 \times 10^{12}$   & $1.16 \times 10^{12}$ &  $6.43 \times 10^{9}$& $1.08 \times 10^{7}$ &$6.44 \times 10^{9}$ \\
    LA  &$1.74 \times 10^{10}$    & $4.68 \times 10^9$ & $2.21 \times 10^{10}$ &  $2.12 \times 10^{11}$& $5.89\times 10^{7}$ & $2.12 \times 10^{11}$ \\
    ZO &$2.01 \times 10^{8}$    & $1.29 \times 10^{10}$ & $1.31 \times 10^{10}$ &  $1.41 \times 10^{7}$& $8.83\times 10^{9}$ & $8.85 \times 10^{9}$ \\
    TO &$6.96\times 10^{10}$ &$1.23 \times 10^{10}$ & $8.19 \times 10^{10}$ &  $7.70 \times 10^{9}$& $6.62 \times 10^{9}$ &$1.43 \times 10^{10}$ \\
    ZO &$6.62 \times 10^{10}$    & $1.62 \times 10^{11}$ & $2.29 \times 10^{11}$ &  $5.66 \times 10^{9}$& $4.75\times 10^{10}$ &$5.32 \times 10^{10}$ \\
       \hline \hline
  \end{tabular*}
  }
\end{table*}

According to Boltzmann transport theory, the carrier mobility can be expressed in Eq. (1) where the electron group velocity and the relaxation time are key parameters. The latter is the inverse of the carrier scattering rate derived from Eq. (2) where the electron, phonon energies, and el-ph coupling matrix elements are needed in a fine $\mathbf{k}$- and $\mathbf{q}$-mesh. The scattering rates at the Dirac points for both stanene and graphene are shown in Table 2 for {\it K} and in Table S1 of the Supporting Information for {\it K \!}$'$. It is found that for graphene, the main contribution of scatterings is the LA mode, whereas the scatterings with ZA mode are negligible. However, for stanene, the scatterings with ZA and TA mode are significantly larger than LA mode. In DPA, only LA phonon scattering is considered. Indeed, the room-temperature mobility of graphene derived from deformation potential theory is in good agreement with the one calculated by full evaluation of el-ph coupling for all phonon modes (Table S1, Supporting Information).\cite{Xi2014Electron-phononApproach} DPA holds in graphene, but not in stanene.

\begin{figure}[htbp]
\centering
 \hbox{\hspace{0.8em}
 \includegraphics[width=\linewidth]{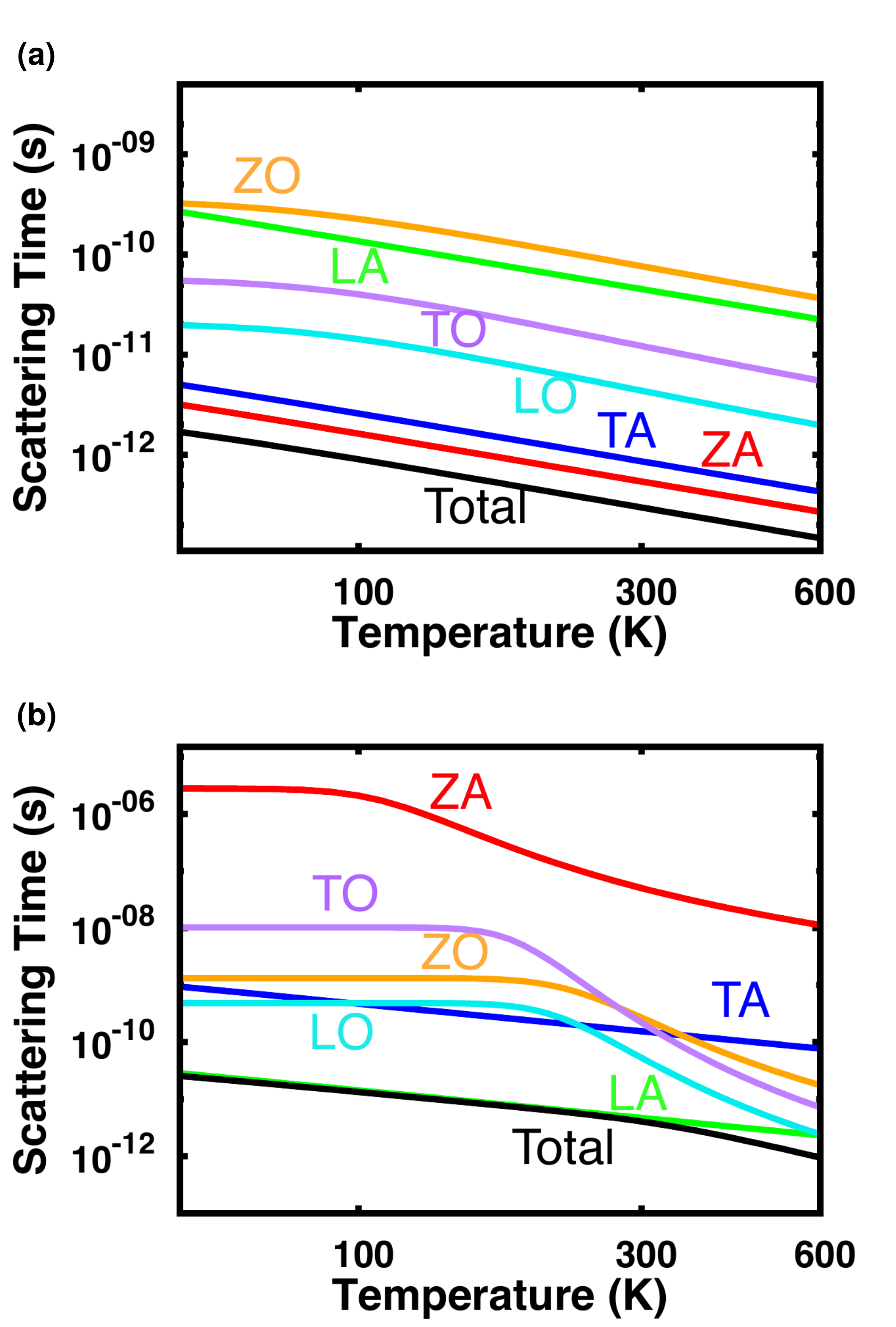}}
   \caption{Temperaturedependenceofscatteringtimefortheconduction band of (a) stanene and (b) graphene for all phonon modes. For graphene, the main contribution to scattering time is the LA phonon. Whereas for stanene, the ZA and TA phonon modes dominate.}
  \label{fgr:el-ph}
\end{figure}

We demonstrate the mode-resolved phonon scattering time as a function of temperature for the sake of comparison between stanene and graphene as in Fig. 3. As expected, for graphene, the curve of ``Total" fully coincides with ``LA". For stanene, however, ``LA" curve is far from the ``Total" meaning that LA contribution is very small. Instead, ``ZA" and ``TA" curves are closer to the ``Total", and even the ``LO" contribution is not negligible. Thus, it is not surprising that the room-temperature mobility of stanene derived from the deformation potential theory is three orders of magnitude larger than that obtained by full evaluation of el-ph coupling for all phonon modes (Table S1, Supporting Information). The mode-resolved scattering rates of the acoustic and optical phonon branch (ZA, TA, LA, ZO, TO, and LO) are separated into intervalley and intravalley scattering as listed in Table 3 for {\it K } and in Table S2 of the Supporting Information for {\it K \!}$'$. For the LA mode, the intravalley scattering overwhelms the intervalley scattering for both graphene and stanene. However, the situation is reversed for TA and ZA modes in stanene, where intervalley scattering overwhelms the intravalley scattering. In graphene, intervalley scattering is the minor contribution among all three acoustic phonon modes. We note that the large intervalley scattering is also reported for silicene.\cite{Gunst2016First-principlesMaterials} Therefore, the intervalley scatterings of TA and ZA for stanene play a dominant role in carrier scattering, and the inadequacy of DPA in stanene is attributed to the overlook of TA and ZA modes. The contribution of ZA phonon scattering can be suppressed through substrate suspension or clamping. Here we also performed the calculation of mobility excluding the ZA phonon. The mobility of stanene was enhanced from 2-3 $\times$ 10$^3$ cm$^2$ V$^{-1}$ s$^{-1}$ to 5-6 $\times$ 10$^3$ cm$^2$ V$^{-1}$ s$^{-1}$ when excluding ZA phonon (Table S3, Supporting Information). 

The temperature dependence of the phonon-limited mobility of stanene is shown in Fig. 4. The ZA and TA modes dominate the optical phonon and LA modes over the whole temperature range. The total mobility obeys a power law of with $\gamma=1.43$. The intrinsic room-temperature mobility of stanene (2-3 $\times$ 10$^3$ cm$^2$ V$^{-1}$ s$^{-1}$) is two orders of magnitude lower that of graphene (2 × 10$^5$ cm$^2$ V$^{-1}$ s$^{-1}$).\cite{Long2009TheoreticalGraphene,Xi2012First-principlesNanomaterials,Xi2014Electron-phononApproach} It is noted that the intrinsic mobility of stanene is smaller compared to graphene, although the weaker el-ph couplings in stanene favor high carrier mobility. However, this is not obscure because the phonon frequency of stanene is one order of magnitude lower than that of graphene. The highest frequency of stanene is smaller than 25 meV, which means that all the phonon modes are excited at room-temperature. Furthermore, the smaller slopes of band structures provide more scattering space in stanene.

\begin{figure}[htbp]
\centering
    \includegraphics[width=\linewidth]{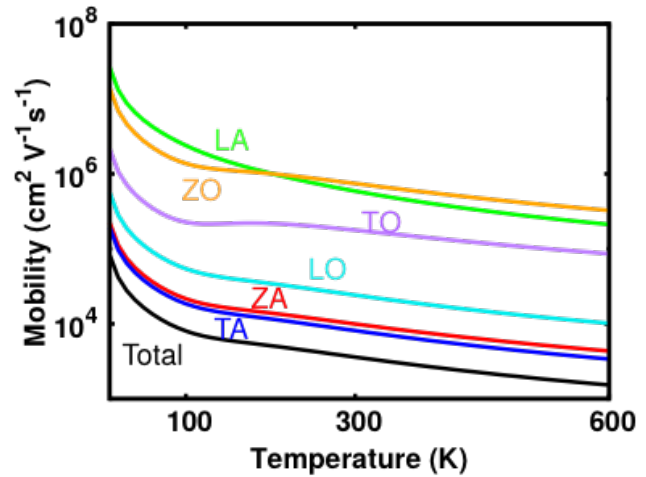}
   \caption{Temperature dependence of intrinsic electron mobility in stanene with phonon mode contribution. }
  \label{fgr:mob-temp}
\end{figure}

  \begin{table*}[htbp]
\small
  \caption{\ Scattering rate of each phonon mode for graphene and stanene at Dirac point  {\it K} and $T=300$. }
  \label{tbl:mode-resolved}
    {\renewcommand{\arraystretch}{1.4}
    \begin{tabular*}{1\textwidth}{@{\extracolsep{\fill}}ccccc}
    \hline \hline
               2D materials  &  $\mu^{DPA}$ (cm$^2$ V$^{-1}$ s$^{-1}$) &$\mu^{EPC}$ (cm$^2$ V$^{-1}$ s$^{-1}$) &Symmetry  & Dominant Phonon  \\
         \hline
    Stanene & 3-4 $\times 10^{6}$   & 2000-3000& D$_{3d}$    &ZA, TA  \\
    Germanene  & $6.2 \times 10^5$\cite{Ye2014IntrinsicCalculations}    & 2800\cite{Gaddemane2016TheoreticalSimulations} & D$_{3d}$ & ZA, TA\cite{Gaddemane2016TheoreticalSimulations} \\
    Silicene   & $2 \times 10^5$\cite{Shao2013First-principlesSilicene} &2100,\cite{Gunst2016First-principlesMaterials} 1200,\cite{Li2013IntrinsicPrinciples} 750\cite{Gaddemane2016TheoreticalSimulations} &  D$_{3d}$ & ZA,\cite{Gunst2016First-principlesMaterials,Gaddemane2016TheoreticalSimulations} TA\cite{Gaddemane2016TheoreticalSimulations} \\
    Graphene & 2-3 $\times 10^{5}$,3  $\times 10^{5}$,\cite{Xi2012First-principlesNanomaterials} 1 $\times 10^{5}$\cite{Hwang2008AcousticGraphene} & 2-3 $\times 10^{5}$,   2 $\times 10^{5}$,\cite{Xi2014Electron-phononApproach} 1.5 $\times 10^{5}$\cite{Gunst2016First-principlesMaterials}    & D$_{6h}$   & LA\cite{Xi2014Electron-phononApproach} \\
    $\alpha$-graphyne& $3 \times 10^{4}$\cite{Chen2013CarrierGraphene.pdf} & $1.6 \times 10^{4}$\cite{Xi2014Electron-phononApproach}  & D$_{2h}$ & LA\cite{Xi2014Electron-phononApproach}\\
   Monolayer MoS$_2$  & 70-200\cite{Cai2014Polarity-reversedNanoribbons} &400,\cite{Gunst2016First-principlesMaterials} 130,\cite{Li2013IntrinsicPrinciples} 410,\cite{Kaasbjerg2012Phonon-limitedPrinciples}　230,\cite{Restrepo2014AMaterials} 150\cite{Li2015ElectricalMoS2}  &  D$_{3h}$   & LA,\cite{Li2013IntrinsicPrinciples,Kaasbjerg2012Phonon-limitedPrinciples} LO$_2$\cite{Kaasbjerg2012Phonon-limitedPrinciples} \\
      \hline \hline
  \end{tabular*}
  }
\end{table*}

\subsection{\!\!\!\!\!Limitation of Deformation Potential Approximation\!\!\!\!\!}
Here we discuss the limitation of DPA by looking at different 2D materials. Table 4 shows the calculated mobilities from DPA ($\mu^{DPA}$) and from full consideration of el-ph coupling ($\mu^{EPC}$) in 2D materials. $\mu^{EPC}$ for stanene, germanene, and silicene, which have bucklings, is several orders of magnitude lower than $\mu^{DPA}$, indicating the discrepancy of DPA. By contrast, DPA is valid for perfectly planar materials such as graphene and $\alpha$-graphyne because $\mu^{DPA} $ and $\mu^{EPC}$ compare well with each other. It seems to imply a concomitant relationship between nonplanarity and the deviation of DPA. For monolayer MoS$_2$, however, even though the sandwiched structure results in nonplanarity, the carriers do not suffer from ZA phonon scatterings.\cite{Gunst2016First-principlesMaterials,Gaddemane2016TheoreticalSimulations,Fischetti2016Mermin-WagnerSymmetry} Indeed, LA phonon scattering dominates in low carrier energy region where carrier excitation energies are lower than optical phonon energies,\cite{Kaasbjerg2012Phonon-limitedPrinciples} and the $\mu^{DPA}$ of MoS$_2$ is comparable with $\mu^{EPC}$. Therefore, nonplanarity does not necessarily invalidate DPA. Actually, for graphene, $\alpha$-graphyne, and monolayer MoS$_2$ (Fig. 5) with $\sigma_h$-symmetry, the intravalley carrier scatterings with the flexural ZA phonon is prohibited according to the Mermin-Wagner theorem.\cite{Fischetti2016Mermin-WagnerSymmetry,Merminf1988EffectStudy} For buckled group IV elemental 2D sheets without $\sigma_h$-symmetry, the intravalley carrier scatterings with the flexural ZA phonons dominate due to the diverging number of ZA phonons.\cite{Fischetti2016Mermin-WagnerSymmetry} In addition to intravalley scattering with flexural ZA mode, our results also point to the limitation of DPA due to the non-negligible intervalley scatterings with ZA and TA. DPA fails when scatterings with ZA or TA dominate.

Based on those facts, we conclude that the limitation of DPA may be caused by (1) flexural ZA phonon scattering in 2D system without $\sigma_h$-symmetry, (2) the intensive intervalley scattering process as seen in stanene. For stanene, as a consequence of buckling, the horizontal mirror symmetry is broken and significant ZA, TA intervalley, and flexural ZA intravalley scattering take place, which results in the breakdown of deformation potential approximation.
\begin{figure}[htbp]
\centering
    \includegraphics[width=\linewidth]{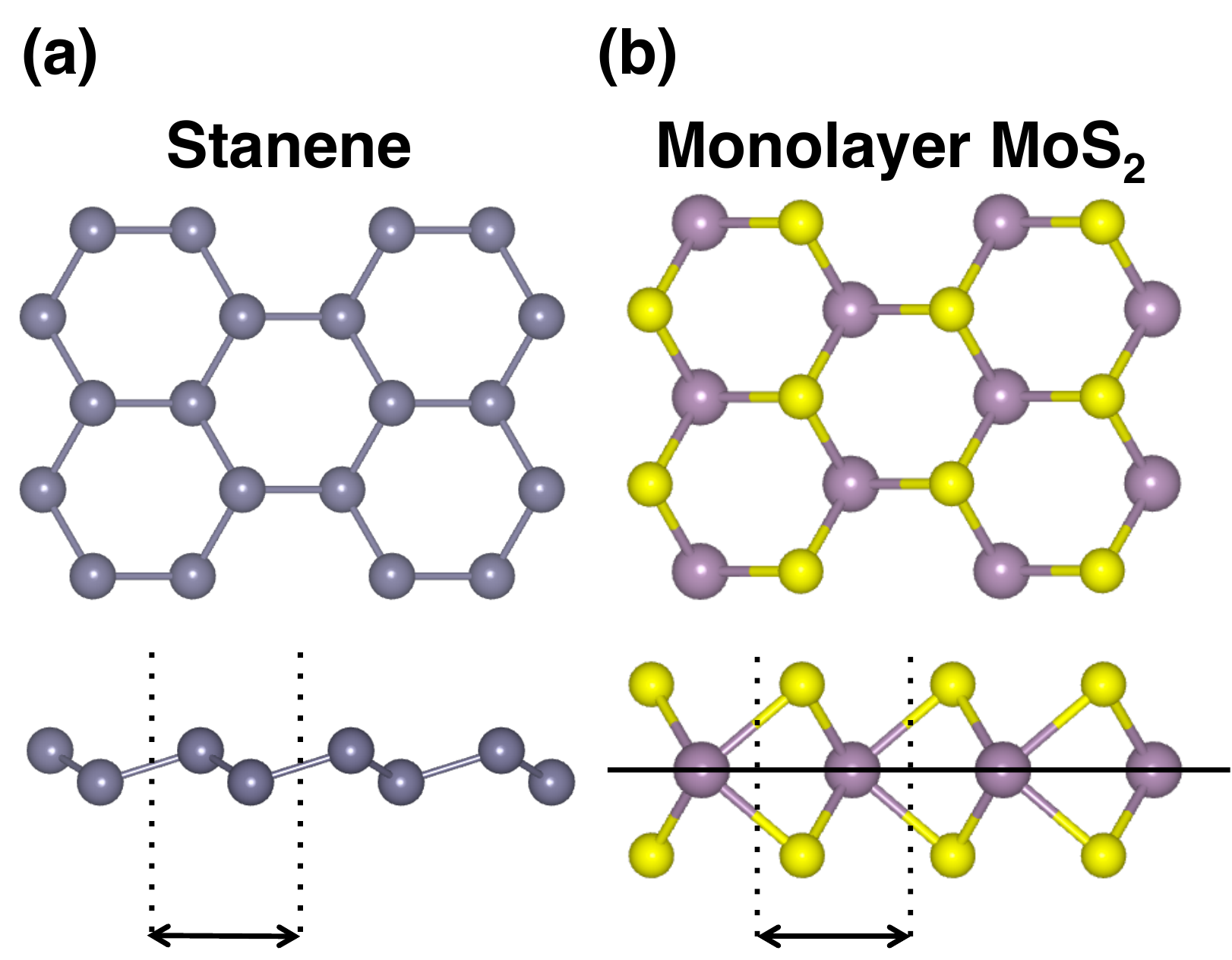}
   \caption{The classification of nonplanarity for (a) stanene, (b) monolayer MoS$_2$. Buckling structure of stanene results in broken horizontal mirror symmetry whereas monolayer MoS$_2$ holds this symmetry due to symmetric sandwiched structure. }
  \label{fgr:mob-temp}
\end{figure}
\section{\label{sec:level1}Conclusion}
To summarize, the phonon-limited charge carrier transport properties of stanene have been studied by performing first-principles DFPT calculations with Wannier interpolation. The intrinsic room-temperature (300 K) mobility is calculated to be 2-3 $\times$ 10$^3$ cm$^2$ V$^{-1}$ s$^{-1}$, which is dominated by the intervalley scattering process of the ZA and the TA phonon modes. The widely employed DPA, assuming the dominant role of LA phonons near the center of the Brillouin zone $(\mathbf{q}=\boldsymbol{\Gamma})$, would overestimate the intrinsic mobility by three orders of magnitude. The invalidity of DPA in stanene is attributed to: (1) the intravalley and intervalley scatterings from the flexural ZA phonon, and (2) the intensive intervalley scattering with TA phonon. DPA has been found to be reasonable for graphene, graphdiyne, graphyne, and TMD such as MoS$_2$ with $\sigma_h$-symmetry. However, it fails in silicene, germanene, and stanene due to lack of $\sigma_h$-symmetry.
We propose to enhance the mobility of stanene through clamping by a substrate in experiment as the mobility increases to 5-6 $\times$ 10$^3$ cm$^2$ V$^{-1}$ s$^{-1}$ when we deliberately excludes the ZA phonon scatterings in our calculation. The intrinsic mobility of stanene is several times larger than that of monolayer MoS$_2$. The high carrier mobility of stanene renders it a promising candidate for nanoelectronic and spintronic applications.

\section*{\label{sec:level1}Appendix}
\subsection{Charge Carrier Mobility and\\ Electron-Phonon Couplings}
In the Boltzmann transport theory, the carrier mobility is expressed
as\cite{Xi2014Electron-phononApproach}
\begin{equation}
\mu =  \frac{\sum_{n, \mathbf{k}} e v^2_{n\mathbf{k}} \tau_{n\mathbf{k}}\left(\frac{\partial f^0_{n\mathbf{k}}}{\partial\varepsilon_{n\mathbf{k}}} \right)}{\sum_{n, \mathbf{k}} f_{n\mathbf{k}}}
\end{equation}
where $\varepsilon_{n\mathbf{k}}$, $v_{n\mathbf{k}}$, $\tau_{n\mathbf{k}}$, and $f^0_{n\mathbf{k}}$ are the electron energy, group velocity,relaxation time, and Fermi-Dirac distribution for the electronic initial state with band index $n$ and wavevector $\mathbf{k}$, and $e$ is the elementary charge. The band index $n$ is summed over conduction (valence) bands for electrons (holes). The group velocity is expressed as $v_{n\mathbf{k}}=\frac{1}{\hbar} \nabla_\mathbf{k}\varepsilon_{n\mathbf{k}}$. The parameter $\tau_{n\mathbf{k}}$ is of prime importance in describing the scattering mechanism of charge carriers. $\tau_{n\mathbf{k}}$ can be given as\cite{Xi2014Electron-phononApproach}
\begin{eqnarray}
\frac{1}{\tau_{n\mathbf{k}}} &=&\frac{2\pi}{\hbar} \sum_{n',\lambda, \mathbf{q}} |g_{\lambda\mathbf{q}}(\mathbf{k},n,n')|^2 \nonumber\\
  &&\times \Large[  ( n_{\mathbf{q}\lambda} + f_{n'\mathbf{k+q}} )
\delta(\varepsilon_{n'\mathbf{k+q}}-\varepsilon_{n\mathbf{k}}-\hbar \omega_{\lambda\mathbf{q}} )   \\
&&+ (n_{\lambda\mathbf{q}} +1 - f_{n'\mathbf{k+q}} ) \delta(\varepsilon_{n'\mathbf{k+q}}-\varepsilon_{n\mathbf{k}}+\hbar \omega_{\lambda\mathbf{q}} )     \Large] \nonumber
\end{eqnarray}
where the sum is performed over all the conduction (valence)
band $n'$ for electrons (holes) and all modes of phonon with mode
index $\lambda$ and wavevector $\mathbf{q}$. $\omega_{\lambda \mathbf{q}}$ and $n^0_{\lambda \mathbf{q}}$ are the frequency and Bose-Einstein distribution of phonons. 
$\varepsilon_{n'\mathbf{k+q}}$ and $f_{n'\mathbf{k+q}}$ represent the electronic energy and the Fermi-Dirac distribution of the final state with band index $n'$ and wavevector $\mathbf{k' = k + q}$. 
\ $g_{\lambda\mathbf{q}}(\mathbf{k},n,n')$ is the key parameter of el-ph coupling matrix element. Within DFPT, $g_{\lambda\mathbf{q}}(\mathbf{k},n,n')$ is calculated as\cite{Baroni2001PhononsTheory}
\begin{equation}
g_{\lambda\mathbf{q}}(\mathbf{k},n,n') =\sqrt{\frac{\hbar}{2M \omega_{{\lambda\mathbf{q}}} }}     \langle \Psi_{n'\mathbf{k+q}}|\frac{\partial V_{KS}} {\partial  \mathbf{u}_{\lambda\mathbf{q}}}\cdot \mathbf{e}_{\lambda\mathbf{q}}|\Psi_{n\mathbf{k}}\rangle 
\end{equation}
where $\frac{\partial V_{KS}} {\partial  \mathbf{u}_{\lambda\mathbf{q}}}$ is the first-order derivative of the self-consistent 
Kohn-Sham potential $V_{KS}$ with respect to the atomic displacement $u_{\lambda\mathbf{q}}$ of phonon mode $\lambda$ and wavevector $\mathbf{q}$, $\mathbf{e}_{\lambda\mathbf{q}}$ is the phonon polarization vector, $|\Psi_{n\mathbf{k}}\rangle$ and $|\Psi_{n'\mathbf{k+q}}\rangle$ are the electronic initial and final Bloch state, respectively, and $M$ is the atomic mass in the unit cell. SOC is not considered in the calculations for phonon dispersion and el-ph couplings.

To calculate the relaxation time and carrier mobility from Eqs. (1) and (2), we need to obtain ultradense $\mathbf{k}$-space electronic band structures, $\mathbf{q}$-space phonon dispersions, and el-ph coupling matrix elements over a fine grid of $\mathbf{k}$- and $\mathbf{q}$-mesh, which is tremendously computationally high cost at the first principles level. However, by implementing the Wannier-Fourier interpolation scheme, accurate electronic energies, phonon frequencies, and el-ph coupling matrix elements can be obtained with reasonable computational burden. Within this scheme, the electron Hamiltonian $H^{el}_{\mathbf{k}}$ with diagonal elements being the eigenstates of electrons, lattice dynamical matrix $D_\mathbf{q}^{ph}$ and el-ph coupling matrix elements $g_{\lambda\mathbf{q}}(\mathbf{k},n,n')$ are first calculated on a coarse $\mathbf{k}$- and $\mathbf{q}$-mesh $N^\mathbf{k(q)}_1 \times N^\mathbf{k(q)}_2 \times N^\mathbf{k(q)}_3$. Second, the electronic Hamiltonian is transformed from Bloch space to Wannier space by using a gauge matrix $\{U_\mathbf{k}\}$ through
 \begin{equation}
H_{\mathbf{R}_e,\mathbf{R}'_e}^{el} = \sum_\mathbf{k} e^{-i\mathbf{k}\cdot (\mathbf{R}'_e-\mathbf{R}_e)} U_\mathbf{k}^\dagger H_{\mathbf{k}}^{el}U_\mathbf{k} 
\end{equation}
where the gauge matrix $U_\mathbf{k}$ is obtained by the maximally localized Wannier functions method.\cite{Marzari2012MaximallyApplications,Souza2001MaximallyBandsb} $R_e$ and $R'_e$ are the lattice vectors in real space, which are also the index of Wannier functions. The lattice dynamical matrix is then Fourier transformed into Wannier space by using the phonon eigenvector matrix $\{\mathbf{e_q}\}$
\begin{equation}
D_{\mathbf{R}_p,\mathbf{R}'_p}^{ph} =\sum_\mathbf{q} e^{-i\mathbf{q}\cdot (\mathbf{R}'_p-\mathbf{R}_p)} \mathbf{e_q} D_{\mathbf{q}}^{ph}\mathbf{e}^\dagger_\mathbf{q}
\end{equation}
where $R_p$ and $R'_p$ are the lattice vectors in real space for phonons. The el-ph coupling matrix is then transformed as\cite{Giustino2007Electron-phononFunctions}
\begin{equation}
\thickmuskip=0mu
\medmuskip=0mu
\thinmuskip=0mu
g(\mathbf{R}_e,\mathbf{R}_q)=\frac{1}{N_p}\sum_{\mathbf{k,q}}e^{-i(\mathbf{k}\cdot \mathbf{R}_e+\mathbf{q}\cdot\mathbf{R}_p)} U_\mathbf{k+q}^\dagger g(\mathbf{k,q}) U_\mathbf{k}\mathbf{e}^{-1}_\mathbf{q}
\end{equation}
where $N_p = N^\mathbf{q}_1 \times N^\mathbf{q}_2 \times N^\mathbf{q}_3$ is the number of unit cells in real space. Herein, we have omitted the band and phonon mode
el index in $g(\mathbf{k,q})$ for simplicity. Since the quantities $H_{\mathbf{R}_e,\mathbf{R}'_e}^{el}$, $D_{\mathbf{R}_p,\mathbf{R}'_p}^{ph}$ and $g(\mathbf{R}_e, \mathbf{R}_q)$ decay rapidly in Wannier space, we can truncate the summation in real space. Finally, we can obtain the electronic energy, phonon frequency, and el-ph coupling matrix elements on a fine $\mathbf{k}$- and $\mathbf{q}$-mesh $N^{'\mathbf{k(q)}}_1 \times N^{'\mathbf{k(q)}}_2 \times N^{'\mathbf{k(q)}}_3$ by diagonalizing the following matrix
\begin{equation}
H_{\mathbf{k'}}^{el} = U_\mathbf{k'}\left(\frac{1}{N_e}\sum_{\mathbf{R}_e}e^{i\mathbf{k'}\cdot \mathbf{R}_e} H_{0,\mathbf{R}_e}^{el}\right) U_\mathbf{k'}^\dagger \\
\end{equation}
\begin{equation}
D_{\mathbf{q'}}^{ph} =\mathbf{e}^\dagger_\mathbf{q'}\left(\frac{1}{N_p}\sum_{\mathbf{R}_q} e^{i\mathbf{q'}\cdot \mathbf{R}_p} D_{0,\mathbf{R}_p}^{ph}\right) \mathbf{e_{q'}} \\
\end{equation}
\begin{equation}
\thickmuskip=0mu
\medmuskip=0mu
\thinmuskip=0mu
\hspace*{-5.7mm} g(\mathbf{k',q'})=\frac{1}{N_e}\sum_{\mathbf{R}_e,\mathbf{R}_q}e^{i(\mathbf{k'}\cdot \mathbf{R}_e+\mathbf{q'}\cdot\mathbf{R}_p)} U_\mathbf{k'+q'}^\dagger g(\mathbf{R}_e,\mathbf{R}_q) U_\mathbf{k'}^\dagger \mathbf{e}_\mathbf{q'}
\end{equation}
where $N_e = N_1^\mathbf{k} \times N_2^\mathbf{k} \times N_3^\mathbf{k}$ is the number of unit cells in real space.

Density functional theory (DFT) and DFPT calcu- lations were performed with Quantum ESPRESSO.\cite{Giannozzi2009QUANTUMMaterials.} Norm-conserving pseudopotential with the Perdew-Burke-Ernzerhof\cite{Perdew1996GeneralizedSimple} exchange-correlation functional was used for the tin atom. For the structure optimization, the plane wave cutoff energy, the convergence threshold, and the force were set to be 65 Ry, 10$^{-12}$ Ry, and 10$^{-6}$ Ry bohr$^{-1}$, respectively. For the phonon dispersion, a $\mathbf{k}$-mesh of 20 $\times$ 20 $\times$ 1 was used for the self-consistent DFT calculation, while the force constants were obtained with a $\mathbf{q}$-mesh of 5 $\times$ 5 $\times$ 1. The Wannier interpolation method was used to obtain ultradense electronic structure, phonon dispersion, and el-ph couplings matrix as implemented in the Wannier90\cite{Mostofi2014AnFunctions} and EPW code.\cite{Noffsinger2010EPW:Functions,Ponce2016EPW:Functionsb} The el-ph coupling matrix was interpolated from 10 × 10 × 1 coarse $\mathbf{k}$-mesh and 5 $\times$ 5 $\times$ 1 coarse $\mathbf{q}$-mesh into 120 $\times$ 120 $\times$ 1 $\mathbf{k}$- and $\mathbf{q}$-meshes for stanene. In graphene, the el-ph coupling matrix was interpolated from 6 $\times$ 6 $\times$ 1 coarse $\mathbf{k}$-mesh and 6 $\times$ 6 $\times$1coarse $\mathbf{q}$-mesh into 120 $\times$ 120 $\times$ 1 $\mathbf{k}$- and $\mathbf{q}$-meshes. Gaussian broadening of 0.045 eV was used for the delta-function in calculating the relaxation time in Eq. (2). We excluded el-ph couplings for phonons with frequency lower than 5 cm$^{-1}$.
\subsection{Deformation Potential Theory}
The el-ph coupling can be given by 
\begin{equation}
g_{\lambda\mathbf{q}}(\mathbf{k},n,n')= \sqrt{\frac{\hbar}{2M \omega_{\lambda\mathbf{q}}}}M_{\lambda\mathbf{q}}
\end{equation}
where the coupling matrix is 
\begin{equation}
M_{\lambda \mathbf{q}}=  \langle \Psi_{n'\mathbf{k+q}}|\frac{\partial V_{KS}}{\partial\mathbf{u}_{\lambda\mathbf{q}}} |\Psi_{n\mathbf{k}}\rangle
\end{equation}
The deformation potential approximation assumes that for long wavelength limit($\mathbf{|q|} \approx 0$), the LA phonon is the dominant scattering process and the coupling matrix could be expressed in terms of the deformation potential constant $D_{LA}$[28]
\begin{equation}
M_{\lambda \mathbf{q}} = D_{LA}|\mathbf{q} |
\end{equation}
Assuming elastic scattering, the approximate relaxation time is then given by
\begin{equation}
\frac{1}{\tau_{n\mathbf{k}}} = \sum_{\mathbf{k'}}\frac{2\pi}{\hbar} \frac{k_B T D^2_{LA}}{C_{2D}}\delta(\varepsilon_{n'\mathbf{k'}}-\varepsilon_{n\mathbf{k}})(1-\cos\theta_{\mathbf{k,k'}})
\end{equation}
When the wavelength of phonon is much larger than the
lattice spacing, the dilation of the unit cell can well reproduce the deformation potential, which is effective potential produced by acoustic wave.

The elastic constant $C_{2D}$ in Eq. (13) is calculated by parabolic curve fitting of the total energy shift $(E- E_0)$ of unit cell  with respect to the lattice strain  $l/l_0$
\begin{equation}
\frac{E-E_0}{S_0}=\frac{C_{2D}}{2}\left(\frac{\Delta l}{l_0}\right)^2 
\end{equation}
where $E_0$ and $S_0$ are the equilibrium total energy and area of the unit cell for 2D materials. The deformation potential constant $D_{LA}$ is obtained by linear regression of the Fermi energy shift  $\varepsilon_F$ with lattice strain  $l/l_0$ for the Dirac cone materials
\begin{equation}
\Delta \varepsilon_F= D_{LA} \left( \frac{\Delta l}{l} \right)
\end{equation}
Computational details can be found in the Supporting Information.
\section*{Aknowledgement}
ZS is deeply in debt to Prof. Daoben Zhu for his insightful discussions, contin- uous encouragements, and fruitful collaborations, which led to more than 30 coauthored publications in the past 15 years. In the occasion of the 75th anniversary, the authors would like to express their sincere wishes to Prof. Daoben Zhu for good health and longevity. This work was supported by the National Natural Science Foundation of China (Grant Nos. 21673123, 21290190, and 91333202) and the Ministry of Science and Technology of China (Grant Nos. 2013CB933503 and 2015CB655002). Computational resources were provided by the Tsinghua University High Performance Supercomputing Center.

\pagebreak

\bibliography{Mendeley.bib}

\pagebreak
\begin{figure*}[htbp]
\thispagestyle{empty}
\centering
    \includegraphics[width=\linewidth]{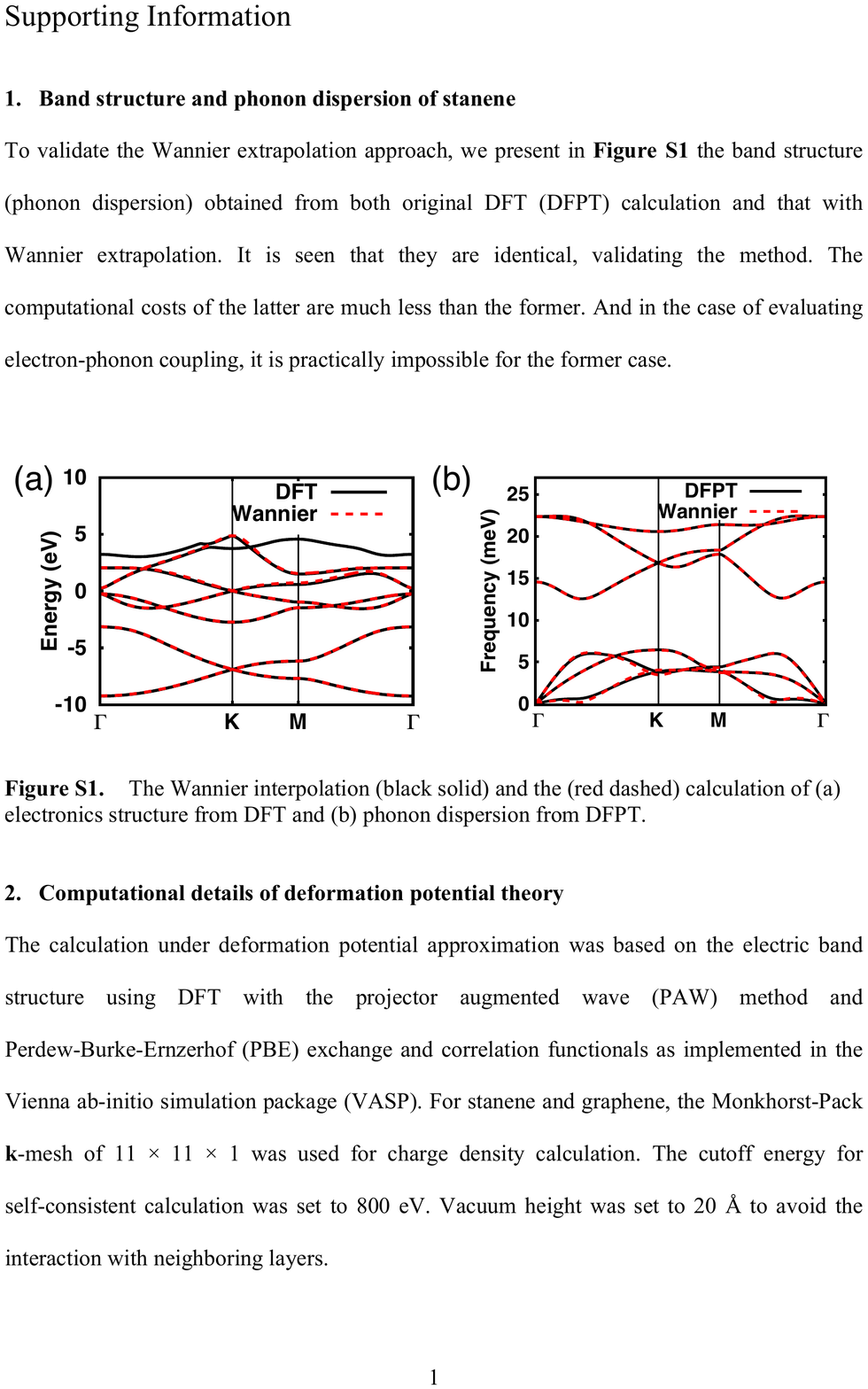}
\end{figure*}

\thispagestyle{empty}
\begin{figure*}[htbp]
\thispagestyle{empty}
\centering
    \includegraphics[width=\linewidth]{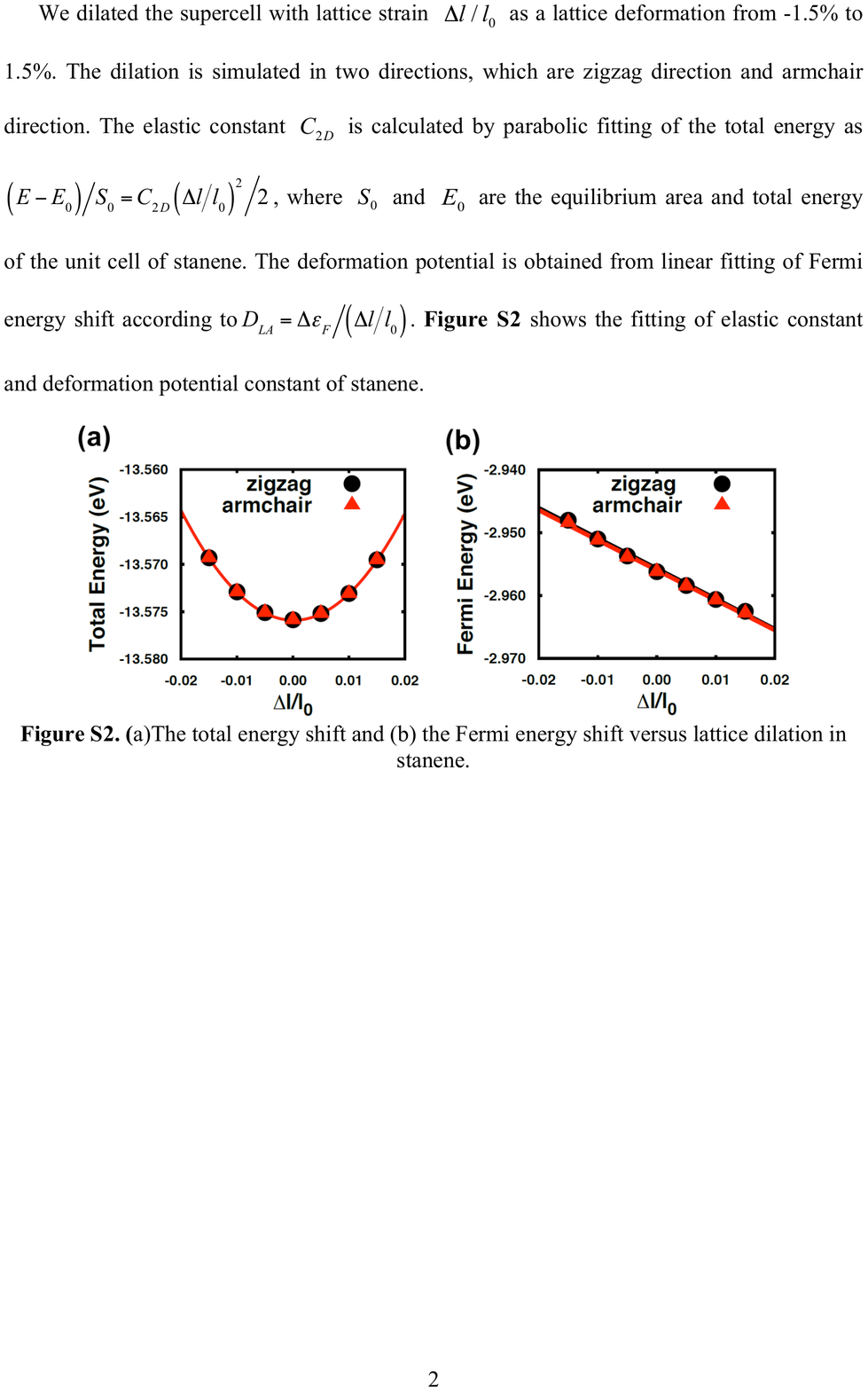}
\end{figure*}
\pagenumbering{gobble}
\begin{figure*}[htbp]
\thispagestyle{empty}
\centering
\pagenumbering{gobble}
    \includegraphics[width=\linewidth]{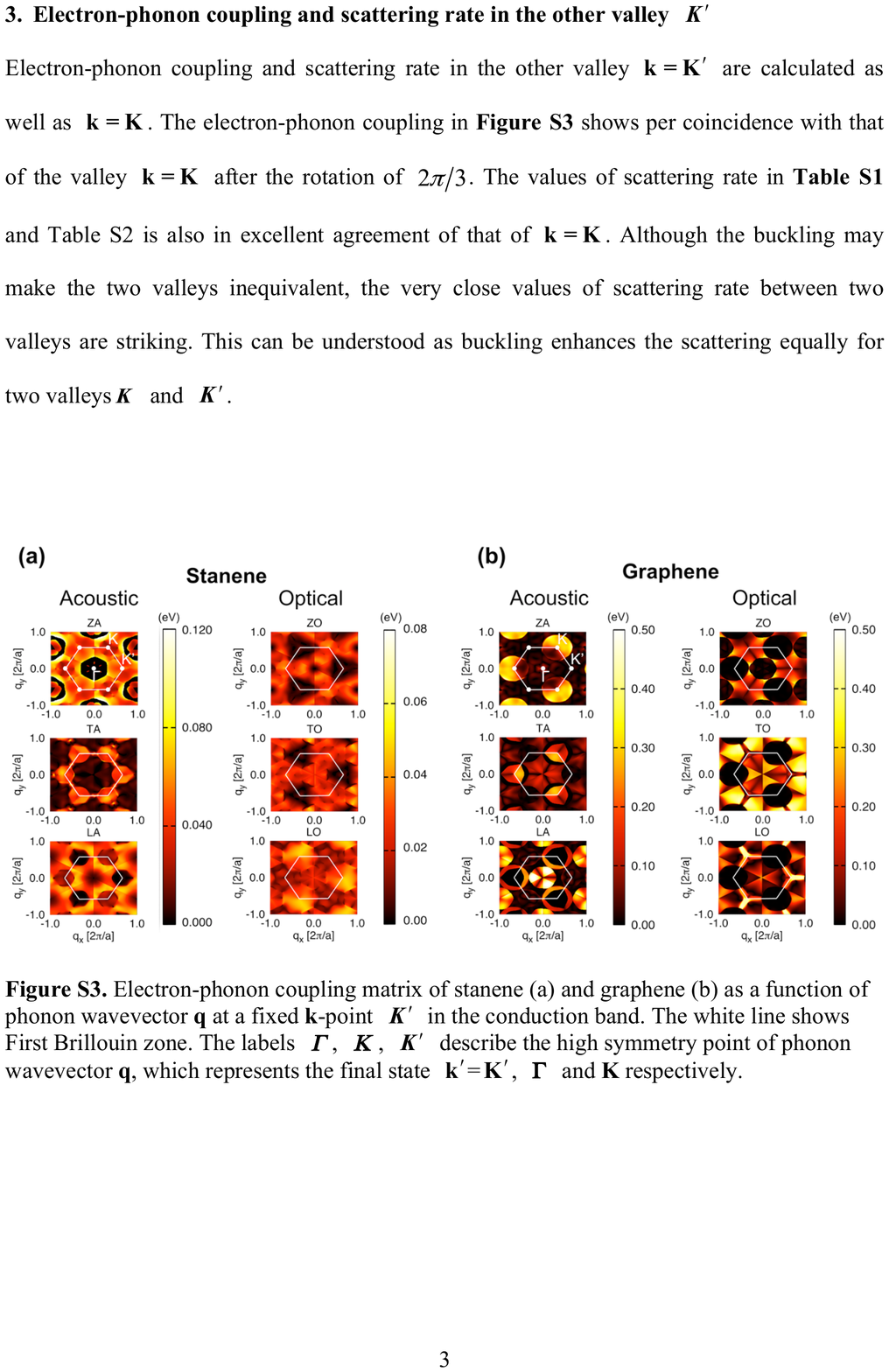}
\thispagestyle{empty}
\end{figure*}

\pagebreak
\begin{figure*}[htbp]
\thispagestyle{empty}
\centering
    \includegraphics[width=\linewidth]{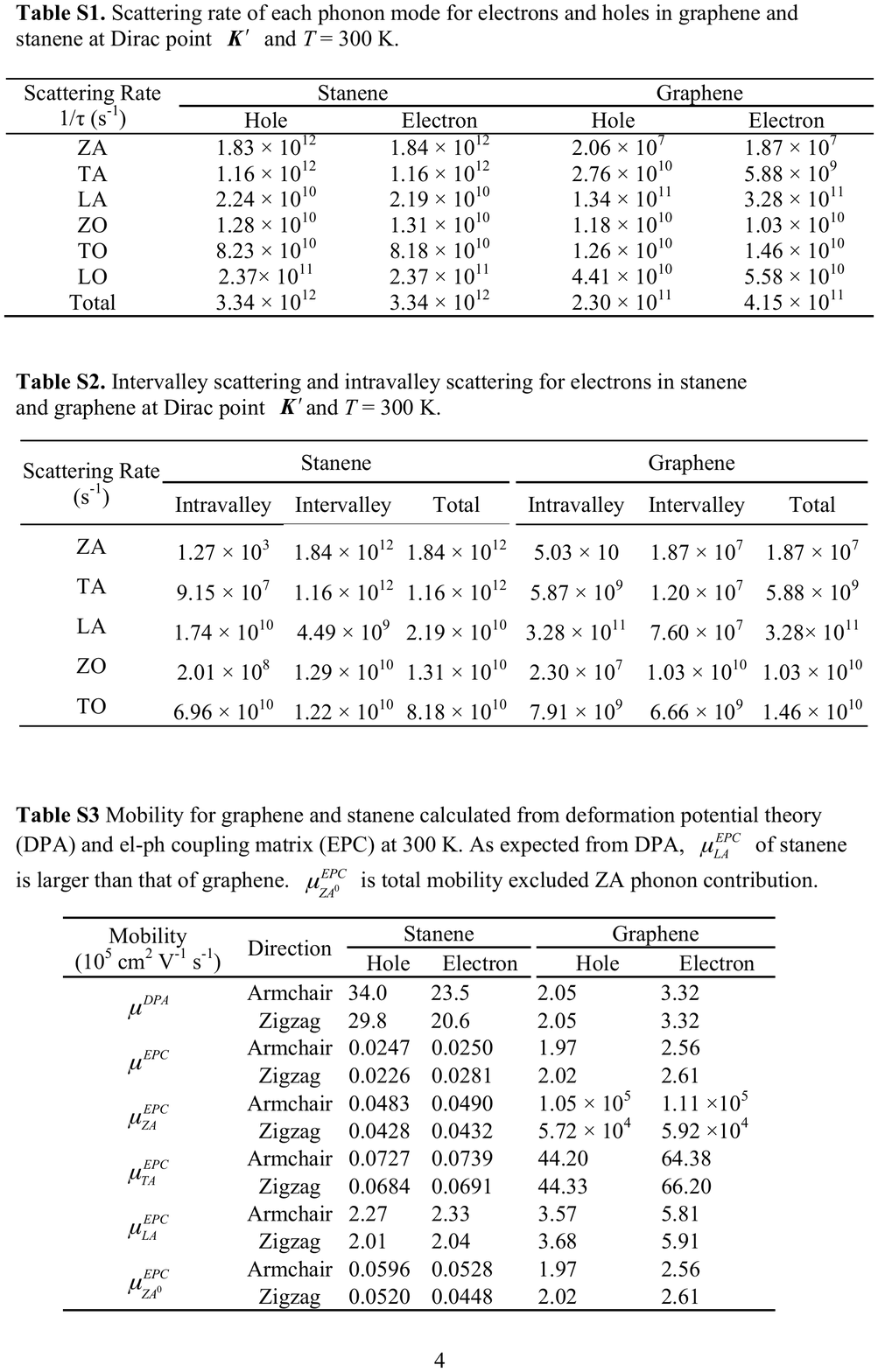}
\end{figure*}
\end{document}